\documentclass{article}

\PassOptionsToPackage{numbers,sort&compress}{natbib}
    \usepackage[final]{neurips_2020}

\usepackage[utf8]{inputenc} 
\usepackage[T1]{fontenc}    
\usepackage[colorlinks = true,
            linkcolor = red,
            urlcolor  = blue,
            citecolor = blue]{hyperref}
\usepackage{url}            
\usepackage{booktabs}       
\usepackage{amsfonts}       
\usepackage{nicefrac}       
\usepackage{microtype}      
\usepackage{algorithm}

\usepackage{graphicx}
\usepackage{amsmath}

\DeclareMathOperator*{\argmax}{arg\,max}
\graphicspath{ {./figures/} }
\usepackage{tikz}

\newcommand\circled[1]{%
  \tikz[baseline=(X.base)] 
    \node (X) [draw, shape=circle, inner sep=-1, fill=black, text=white] {\strut \footnotesize #1};%
}

\title{Factorized Neural Processes for Neural Processes: $K$-Shot Prediction of Neural Responses}

\author{{R. James Cotton,\textsuperscript{1,a}
Fabian H. Sinz,\textsuperscript{2-5,b,*}
Andreas S. Tolias,\textsuperscript{4-5,c,*}}\\
\\ \textsuperscript{1} Department of PM\&R, Northwestern University, Shirley Ryan Ability Lab\\
\textsuperscript{2} Institute for Bioinformatics and Medical Informatics, University Tübingen, Germany\\
\textsuperscript{3} Bernstein Center for Computational Neuroscience, University of Tübingen, Germany
\\ \textsuperscript{4} Department for Neuroscience, Baylor College of Medicine, Houston, TX, USA
\\ \textsuperscript{5} Center for Neuroscience and Artificial Intelligence, Baylor College of Medicine, Houston, TX, USA
\\\\\textsuperscript{a}\texttt{rcotton@sralab.org}, \textsuperscript{b}\texttt{fabian.sinz@uni-tuebingen.de}, \textsuperscript{c}\texttt{astolias@bcm.edu}
\\ \textsuperscript{*}equal contribution
}

\begin{document}

\maketitle

\begin{abstract}
In recent years, artificial neural networks have achieved state-of-the-art performance for predicting the responses of neurons in the visual cortex to natural stimuli. However, they require a time consuming parameter optimization process for accurately modeling the tuning function of newly observed neurons, which prohibits many applications including real-time, closed-loop experiments.
We overcome this limitation by formulating the problem as $K$-shot prediction to directly infer a neuron's tuning function from a small set of stimulus-response pairs using a Neural Process. 
This required us to developed a Factorized Neural Process, which embeds the observed set into a latent space partitioned into the receptive field location and the tuning function properties.  
We show on simulated responses that the predictions and reconstructed receptive fields from the Factorized Neural Process approach ground truth with increasing number of trials. 
Critically, the latent representation that summarizes the tuning function of a neuron is inferred in a quick, single forward pass through the network.
Finally, we validate this approach on real neural data from visual cortex and find that the predictive accuracy is comparable to --- and for small $K$ even greater than --- optimization based approaches, while being substantially faster.
We believe this novel deep learning systems identification framework will facilitate better real-time integration of artificial neural network modeling into neuroscience experiments.

\end{abstract}

\section{Introduction}
There is a long and rich history of modeling the response of visual cortex neurons to stimuli extending back to the work of Hubel and Wiesel on simple and complex cells \cite{hubel1962receptive}. 
In recent years, artificial neural networks (ANNs) have achieved state-of-the-art performance predicting neural responses to natural stimuli \cite{Antolik2016,Yamins2016,Klindt2017,Cadena2019,Batty2016,Sinz2018,Yamins2014,Vintch2015,Ecker2018,Walker2019}. 
These models are accurate enough that the stimuli that maximally excite a neuron can be computed \textit{in silico}, and when tested \textit{in vivo} indeed drive neurons effectively \cite{Walker2019,Bashivan2019}.
However, these approaches place the computational burden of optimizing network parameters \textit{after} extensive data from a neuron has been collected, which prohibits their use in real-time closed-loop experiments.
To avoid this optimization step, we wanted a model that can predict the response of a novel neuron to any stimulus, conditioned on a set of $K$ observed stimulus-response pairs -- essentially performing $K$-Shot prediction on neural responses.

\citet{Garnelo2018a} aptly describe how Neural Processes (NPs) can help solve this problem:
\emph{``Meta-learning models share the fundamental motivations of NPs as they shift workload from training time to test time. NPs can therefore be described as meta-learning algorithms for few-shot function regression''}. 
NPs achieve this by embedding input and output measurements into a latent space that maps to a space of functions, essentially learning the distribution over functions and a method to infer the posterior over functions given limited samples \cite{Garnelo2018a,Garnelo2018,Kim2019}. 

A significant advance in modeling visual responses with ANNs was using convolutional neural networks with a factorized readout between the tuning function's location and properties  \cite{Klindt2017, Sinz2018}.
We found that NPs struggle to learn the space of tuning functions from stimulus-response samples without such a factorized representation. 
Thus, we developed a Factorized Neural Process (FNP), which is composed of stacking multiple NPs.
A key insight for this was that by passing the latent variable computed by early layers to deeper layers, in addition to the observations, we could obtain a factorized latent space while retaining the representational power and efficiency of NPs. 
We used a two-layer FNP applied to visual responses, where the first NP produces a latent variable for the tuning function's location that the second NP uses to infer the tuning function's properties.
We found that a FNP trained on simulated data generalizes to new neurons, successfully inferring the tuning function's location and properties and predicting the responses to unseen stimuli. An FNP trained on neural responses from the mouse primary visual cortex made predictions with comparable accuracy to state-of-the-art approaches, and made these predictions almost \emph{100 times faster}.

In short, our contributions in this work include: \circled{1}  We reformulate the problem of predicting the response of neurons to visual stimuli as a K-shot regression problem, removing the time consuming step of optimizing network parameters for each newly acquired neuron. \circled{2} We develop a Factorized Neural Process that embeds the observed stimuli-response pairs into a latent space representing the tuning function that is partitioned into location and tuning function properties. \circled{3} We train this Factorized Neural Process for Neural Processes end-to-end on simulated data and show it approaches the ground truth predictions as the size of the observation set increases. \circled{4} We found that this approach performs comparably to state-of-the-art predictive models on responses from mouse visual cortex neurons while improving estimation speed by multiple orders of magnitude. The code is available at \url{https://github.com/peabody124/fnp_neurips2020}.

\section{Neural Processes for Neural Processes}

The core steps that allow a NP to efficiently meta-learn a $K$-shot regression model are (1) encoding each element from a set of observed input-output observations into a representation space, (2) aggregating the elements in that representation space (typically by taking the mean) to produce a sufficient statistic of the observed set, (3) a conditional decoder that maps the aggregated representation to a function used for prediction of new observations, and (4) training this over many different sets of observations from different sample functions, i.e. meta-learning the distribution over tuning functions \cite{Garnelo2018}. Our approach is largely based on \citet{Garnelo2018a}, which expanded on \citet{Garnelo2018} by introducing a stochastic variable used by the conditional decoder. NPs were further extended to include attention in \citet{Kim2019}, which we do not use in this work. 

First, we describe the data generation process we seek to model: Let $\mathcal F : \mathcal X \rightarrow \mathcal Y$ be the space of all tuning functions that map from images to neural responses. An individual neuron corresponds to a sample function, $f \in \mathcal F$, from which we get $K$ observations $O_K=\{(\boldsymbol x_i, y_i)\}_{i=0}^{i<K}$ where $y_i \sim f(\boldsymbol x_i)$,  $\boldsymbol x \in \mathcal X$ and $y \in \mathcal Y$. The model should maximize the likelihood of a new input-output observation, called the target: $p_\theta(y_t | \boldsymbol x_t, O_K)$ with $y_t \sim f(\boldsymbol x_t)$, i.e. $K$-shot prediction.  

Following the formulation of stochastic NPs \cite{Garnelo2018a}, this predictions is split into \emph{encoding} the set of observations into a posterior distribution over a latent space, $p_\theta(\boldsymbol z|O_K)$ with $\boldsymbol z \in \mathbb R^D$, and \emph{conditionally decoding} it with a predictive distribution conditioned on the latent variable and target input, $p_\theta(y_t|\boldsymbol x_t, \boldsymbol z)$:
\begin{equation}
p_\theta(y_t | \boldsymbol x_t, O_K) =\int p_\theta(y_t|\boldsymbol x_t, \boldsymbol z)\, p_\theta(\boldsymbol z|O_K) \, \mathrm d \boldsymbol z
\label{eq:np_latent}
\end{equation}
The encoder computes the distribution over $\boldsymbol z$ using a learned embedding of individual observations that are aggregated into a sufficient statistic: $s_K=\frac{1}{K}\sum_{i<K} h_\theta(\mathbf x_i, y_i)$, with $h_\theta: \mathcal X \times \mathcal Y \rightarrow \mathbb R^{D_2}$ implemented as a neural network. The dimensionality of this statistic, $D_2$, does not need to match the dimensionality of the latent variable, $D$, but it is used to parameterize a distribution for the latent variable, such as Gaussian: $p_\theta(\boldsymbol z | O_K) \sim \mathcal N(\mu_\theta(s_K), \sigma_\theta(s_K))$. The decoder function is also implemented as a distribution that is parameterized by the output of a neural network, in our case determining the mean rate of a Poisson distribution $\lambda_\theta : \mathcal X \times \mathbb R^D \rightarrow \mathbb R$ and $p_\theta(y_t|\boldsymbol x_t, \boldsymbol z) \sim \mathtt {Poisson}[\lambda_\theta (\boldsymbol x_t, \boldsymbol z)]$.

Our formulation differs from \cite{Garnelo2018, Kim2019} in a few details. First, they used a variational approximation to train the encoder. Because of our approach to efficient K-Shot training, described below in Section~\ref{section:k_shot_training}, that involves predicting single samples we can directly optimize Eq.~\ref{eq:np_latent} with samples of $\boldsymbol z$. Second, and more critically, for NPs applied to image completion, individual observations are pixels with $\mathcal X \in \mathbb R^2$ and $h_\theta$ was implemented with an MLP \cite{Garnelo2018,Garnelo2018a, Kim2019}. To apply this to direct regression of neural responses, $h_\theta$ must embed entire stimulus-response associations where the domain of $\mathcal X$ are images $\mathbb R^{\mathrm{width} \times \mathrm{height}}$.

We attempted to apply a NP to predicting visual neuron response by including a convolutional neural network (CNN) in $h_\theta$ to extract the relative stimulus features and optimizing Eq.~\ref{eq:np_latent}. 
However, we could not achieve good predictive performance despite trying numerous architecture alterations. 
This motivated us to develop the Factorized Neural Process.

\subsection{Factorized Neural Process}

\begin{figure}
  \centering
  \includegraphics[width=0.8\linewidth]{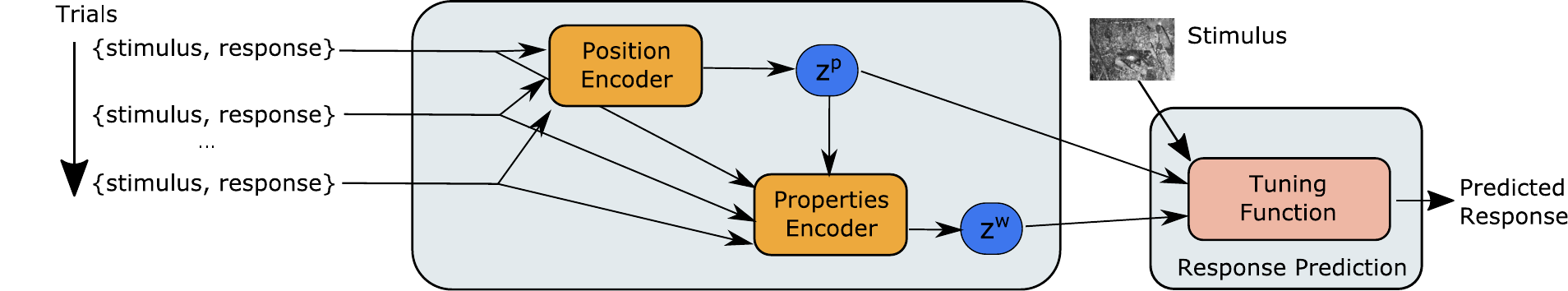}
  \caption{Overview of K-shot neural prediction using a Factorized Neural Process.}
  \label{fig:factorized_latent_process}
\end{figure}

Prior work using neural networks to model visual responses has demonstrated the importance of identifying the location of a receptive field before trying to model the exact properties. For example, early work used a two stage approach which first computed a spike-triggered average to localize the receptive fields and then stimuli were cropped and centered on this location to build predictive models \cite{Batty2016}. In Section~\ref{section:tuning_position} we make a link between a spike-triggered average and our position encoder. 
When using CNNs to predict neural responses, \citet{Klindt2017} showed it is helpful to factorize the tuning function into a location and the tuning properties within that location. We used a similar approach to factor our latent space into visual location and response properties.

Aggregation in NPs should be invariant to permutation of the order of elements in the set, because changing the order inputs are presented to the function should not change the representation of that function (at least under the generative model of our data stated above). Aggregation using the mean, such as used in the NPs above, is one aggregation operation with this property. \citet{Zaheer2017} demonstrated that it applies to a class of deeper network architectures, which they termed Deep Sets. These assure permutation invariance when composing multiple operations, provided each operation (layer) transforms and passes individual elements from a set along with the result of an invariant aggregation operation (such as mean). This insight encouraged us to develop a FNP by composing multiple NPs. 

In a FNP, the encoder, $h_\theta^l$, for each layer $l$ receives both observed samples and the latent variable samples from lower layers, $z^l$ (which is an empty set for the first layer):
\begin{equation}
    s_K^l :=  \frac{1}{K}\sum_{i<K} h^l_\theta(\boldsymbol x_i, y_i, \left\{ \boldsymbol z_K^j \right\}^{j<l}), \qquad
    \boldsymbol z_K^l \sim p_\theta^l \left(\boldsymbol z_K^l | s_K^l\right)
    \label{eq:s_k_l}
\end{equation}
The decoder receives all the latent variables, which can be marginalized out to give the  predictive model for a $L$ layer FNP:
\begin{equation}
    p_\theta(y_t | \boldsymbol x_t, O_K) = \int _{\boldsymbol z} p_\theta \left( y_t | \boldsymbol x_t , \left\{ \boldsymbol  z_K^j \right\} \right) \prod_{l=0}^{l<L} p_\theta^l \left(\boldsymbol z^l| s_K^l\right) \mathrm{d} \boldsymbol z_K^0 ... \mathrm{d} \boldsymbol z_K^{L-1}
    \label{eq:fnp}
\end{equation}

\subsection{FNP Model for Visual Responses}

Following the motivation above, we use a two-layer FNP to partition the latent space into the tuning function position, $\boldsymbol z_K^p$, and the tuning function properties within that location, $\boldsymbol z_K^w$ (Figure \ref{fig:factorized_latent_process}). The factorized probability (following Eq.~\ref{eq:fnp}) is thus:
\begin{equation}
p_\theta(y_t | \boldsymbol x_t, O_K) = \int _{\boldsymbol z_K^p, \boldsymbol z_K^w} p_\theta\left(y_t | \boldsymbol x_t , \boldsymbol  z_K^w, \boldsymbol z_K^p\right) p_\theta\left(\boldsymbol z_K^w | s^w_K\right) \, p_\theta \left( \boldsymbol z_K^p | s^p_K \right)  \mathrm{d} \boldsymbol z_K^p \mathrm{d} \boldsymbol z_K^w \\
\label{eq:vision_fnp}
\end{equation}
The latent space acquires this factorization from the inductive bias in the architecture of the conditional decoder (Section~\ref{section:response_predictor}), which uses a similar factorization between tuning position and properties as \citet{Klindt2017}. With this conditional decoder, we empirically found certain architectures helped the encoders learn an appropriate embedding for these latent variables, which we describe in Sections \ref{section:tuning_position} and \ref{section:properties_encoding}. All components use a common Group Convolutional Neural Network to compute the relevant features space from the stimuli, which we will describe next. 

\subsection{Group Convolutional Network}

One of the most striking features of V1 neurons are their sensitivity to stimulus orientation \cite{hubel1962receptive}, so CNN-based models expend a great deal of capacity representing the same feature at different orientations. \citet{Cohen2016, Cohen2016a} developed  Group Convolutional Neural Networks (G-CNNs), which are rotationally equivariant network that reuse parameters across orientations, similarly to how CNNs reuse parameters across space. G-CNNs have improved performance on a number of image classifications tasks \cite{Weiler2017,Bekkers2018,Lafarge2020} including the modeling of visual cortex neurons \cite{Ecker2019,Ustyuzhaninov2019}. They have also been combined with a DenseNet \cite{Huang2017} architecture when modeling histology images \cite{Veeling2018}. Our stimuli were grayscale images, $\mathcal X \in \mathbb R ^{H_{\boldsymbol x} \times W_{\boldsymbol x} }$ that were transformed to a reduced spatial dimension with multiple feature maps: $g_\theta:\mathbb R^{H_{\boldsymbol x} \times W_{\boldsymbol x} } \rightarrow \mathbb R^{H \times W \times C} $.

Our G-CNN Dense Blocks followed the bottleneck DenseBlock architecture of \cite{Huang2017} (other than using group convolutions) of 1) Batch Normalization \cite{Ioffe2015}, 2) non-linearity, 3)  $1\!\times\! 1$ group convolution, 3) Batch Normalization, 4) non-linearity, 5) spatial group convolution, 6) concatenating this output to the inputs. Specific architectural parameters are described in the experiments. Following \citet{Bekkers2018} and \citet{Veeling2018}, images were passed through an initial layer that lifted them into a group representation, followed by the G-CNN Dense Blocks, and a subsequent projection from the group representation (i.e. a tensor with an additional dimension representing orientation) down to a representation over image space by flattening the group and channel dimensions together. An additional $1\!\times\! 1$ convolution \cite{Szegedy2015,Lin2013} reduce this to the final channel depth. We found using the G-CNN DenseNet architecture resulted in faster and more reliable training than a standard CNN, although we did not perform exhaustive searches over architectures. 

\subsection{Tuning function position encoder}
\label{section:tuning_position}

The first layer is designed to produce a distribution for a latent variable centered at the location of the tuning function, $\boldsymbol z_K^p$. The latent has two dimensions corresponding to horizontal and vertical position.
To compute this, we found it was important to preserve spatial structure from $g_\theta(\boldsymbol x)$ when performing the set aggregation, which we did by averaging the embedded observations into an image analogous to a trainable, non-linear, spike-triggered average:
\begin{equation*}
    s_K^p =  \frac{1}{K} \sum_{i<K} h^p_\theta \left( \left[ g_\theta(\boldsymbol  x_i) ,  y_i g_\theta(\boldsymbol  x_i) , y_i  \right] \right), \qquad s_K^p \in \mathbb R^{H \times W}
\end{equation*}
Where $[\cdot,\cdot]$ concatenates on the channel dimension (and tiles non-spatial values, such as $y_i$, along the spatial dimension) and  $h^p_\theta$ is a series of $1\!\times\! 1$ convolutions that outputs a single channel; the first two $1\!\times\! 1$ retain the same channel depth as $g_\theta$ followed by an ELU activation \cite{Clevert2015}. We interpret $\mbox{softmax}(s_K^p)$ as a spatial distribution over the receptive field location of a given neuron. In order to sample a single spatial readout point, we separately computed the marginal means $\boldsymbol\mu(s_K^p)$ and standard deviations $\boldsymbol\sigma(s_K^p)$ of $\mbox{softmax}(s_K^p)$, and used them to parameterize a truncated normal distribution $\mathcal N_{\mid}$, cut at the same height and width as the output of the G-CNN ($H \times W$): 
\begin{equation*}
    \boldsymbol  z_K^p \sim \mathcal N_{\mid}\left(\boldsymbol \mu(s_K^p), \boldsymbol \sigma(s_K^p) \right) , \qquad \boldsymbol z_K^p, \boldsymbol \mu(s_K^p), \boldsymbol \sigma(s_K^p) \in \mathbb R^2.
\end{equation*}
We also experimented with learning an MLP to transform $s_K^p \rightarrow (\boldsymbol \mu, \boldsymbol \sigma)$. While this worked, it was much slower to train than the Gaussian approximation of $\mbox{softmax}(s_K^p)$ we used.

\subsection{Tuning function properties encoder}
\label{section:properties_encoding}
The second layer of the FNP computes the tuning function's properties at the location determined by the first layer:
\begin{equation}
    s_K^w  = \frac{1}{K} \sum_{i<K} h^w_\theta \left( \left [\mathcal T \left( g_\theta(\boldsymbol  x_i\right), \boldsymbol z_K^p) , y_i \right] \right),
    \label{eq:tuning_position}
\end{equation}
where $\mathcal T: \mathbb{R}^{H \times W \times C} \times \mathbb R^2 \rightarrow \mathbb{R}^C$ is a differentiable operator that performs interpolation on the image features at a particular location to preserve differentiability with respect to the spatial location. We use a spatial transformer layer for that \citep{jaderberg2016spatial}. 
Essentially, \eqref{eq:tuning_position} selects the features computed by the G-CNN at a given location, concatenates them with the response, and passes them through a two-layer MLP, $h_\theta^w$.

We split $s_K^w=(\boldsymbol \mu_K^w, \boldsymbol \Sigma_K^w)$ into a mean and covariance (more specifically the vectorized lower triangular component of the covariance) and use it to parameterize the distribution over the $D$-dimension latent variable for tuning function's properties within that location:
\begin{equation*}
    \boldsymbol z_K^w \sim \mathcal N(\boldsymbol \mu_K^w, \boldsymbol \Sigma_K^w) , \qquad \boldsymbol z_K^w, \boldsymbol \mu_K \in \mathbb R^D.
\end{equation*}

\subsection{Conditional Decoder}
\label{section:response_predictor}
The latent variables for tuning function position and tuning function properties, $\boldsymbol z_K^p$, $\boldsymbol z_K^w$, summarize the observed set of stimulus-response pairs, $O_K$. They define a $K$-shot regression tuning function that we use them to predict the response to new target stimuli, $\boldsymbol x_t$.
\begin{equation*}
    \lambda(\boldsymbol x_t, \boldsymbol z_K^p, \boldsymbol z_K^w) := \mathtt{ELU}\left(\left[\mathcal T \left( g_\theta\left(\boldsymbol  x_t\right), \boldsymbol z_K^p\right), 1\right] \cdot u_\theta\left(\boldsymbol z_K^w\right) \right) + 1
\end{equation*}
Similar to prior work \cite{Klindt2017, Ecker2019, Walker2019,Sinz2018,Cadena2019}, our response predictor uses a linear weighting of the features from the convolutional network selected at the neuron's tuning function's location. Because we sometimes used a different dimensional representation for tuning function's properties than the feature map, we used an additional two-layer MLP, $u_\theta$ to match the dimensions. We also found this additional transformation of the latent improved performance, although did not experiment with this extensively. The predictive response probabilities for $y_t$ are modeled as a Poisson distribution with mean rate $\lambda$:
\begin{equation}
p_\theta(y_t|\boldsymbol x_t, \boldsymbol z_K^p, \boldsymbol z_K^w) \sim \mathtt {Poisson}[\lambda(\boldsymbol x_t, \boldsymbol z_K^p, \boldsymbol z_K^w)]
\label{eq:conditional_decoder}
\end{equation}

\subsection{Efficient K-Shot Training}
\label{section:k_shot_training}
Each training sample for this model consists of a set of K stimuli and responses with an additional target pair and it must be trained over many sets. If implemented na\"ively and sampling a range of set sizes, training is computationally prohibitive as it requires passing thousands of images through the G-CNN per sample. Additionally, in this case the loss is determined by the prediction of a single sample, thus the gradient is fairly noisy. To greatly accelerate training we used two techniques.

The first technique was including the responses of multiple neurons to the same stimuli in a minibatch. This allowed the computation of $g_\theta(\boldsymbol x)$ to be reused across the neurons, at the expense of GPU memory. This memory is proportional to the number of stimuli $\times$ the number of neurons, because Eq.~\ref{eq:tuning_position} is computed in parallel for all neurons and must be stored during each step to allow backpropagation through the G-CNN. 

The second technique was for a set of $T$ trials, we computed the $K$-shot prediction for every set size up to $T-1$ by predicting the response on the next trial. Because aggregation involved computing the mean outputs along the set dimension, this could be efficiently implemented with a cumulative sum that excluded the current element divided by the set size. This allowed us to test $T-1$ predictions rather than 1 for each training set with almost no increase in computation. Thus the per-neuron loss function we minimized was:
\begin{equation*}
    \mathcal L(\theta) = -\sum_{k=0}^{T-1} \log p(y_{k+1} | \boldsymbol x_{k+1}, O_k) 
\end{equation*}
With the likelihood approximated by the FNP described in Eq. \ref{eq:vision_fnp}.  We do not compute the full marginalization in Eq.~\ref{eq:vision_fnp} but estimate it using Monte Carlo samples from $\boldsymbol z_k^w$ and $\boldsymbol z_k^p$ drawn for each value of $k$. The distribution parameters are optimized using the reparameterization trick \cite{Kingma2014}.

In prior work with NPs (that had much less computationally expensive inputs), the full $T$ trials were randomly partitioned into the observation and target set to learn generalization over set size \cite{Garnelo2018, Garnelo2018a, Kim2019}. We leave it to future work to determine if this can be efficiently done with our model, for example generating multiple splits for the encoded values in a single batch, and if it improves training speed or final performance.

\section{Experiments with Simulated Neural Responses}

We first validated that using our approach, a trained FNP could infer the ground truth tuning function from simulated simple and complex visual neurons. Experiments using different architectures and hyperparameters were managed using DataJoint \cite{Yatsenko2015}. Please see the Appendix for the architecture details used in the presented results.

\subsection{Simulated responses}
\label{section:simulated_responses}
Simulated neurons had a linear component to their receptive field that was generated using a Gabor kernel, $k_\phi$, with parameters $\phi$ including $x$ and $y$ location, orientation, frequency, width, phase offset and scale. Cells could either be simple or complex. Responses for simple cells were generated with the inner product between the receptive field kernel and the stimuli, followed by a ReLU non-linearity. Complex cells used an energy-based model which involved a second kernel, $k_{\phi+}$, with the same parameters as the first but with the phase offset by $\pi/2$. Responses were then sampled from a Poisson distribution for a given cell. 
\begin{equation*}
    \lambda_\phi(\boldsymbol x) = 
    \begin{cases}
    \mathtt{ReLU} \left( \boldsymbol x \cdot k_\phi \right)  & \text{, if $\phi$ simple} \\
    \sqrt{ \left( \boldsymbol x \cdot k_\phi \right)^2 + \left( \boldsymbol x \cdot k_{\phi+} \right)^2} & \text{, if $\phi$ complex}.
    \end{cases},
    \qquad r_i \ \sim \mathtt{Poisson}[\lambda_\phi(\boldsymbol x_i)]
\end{equation*}

To simulate the finite data available in real experiments, we used a fixed set of 5000 neuron parameters and 10,000 stimuli in the training data and a different set in the validation data. Stimuli were sampled from ImageNet downsampled to $16 \times 16$ or $32 \times 32$ \cite{Russakovsky2015,Chrabaszcz2017}. Each sample used a random stimulus order with a fixed number of trials.

\begin{figure}
  \centering
  \includegraphics[width=0.5\linewidth]{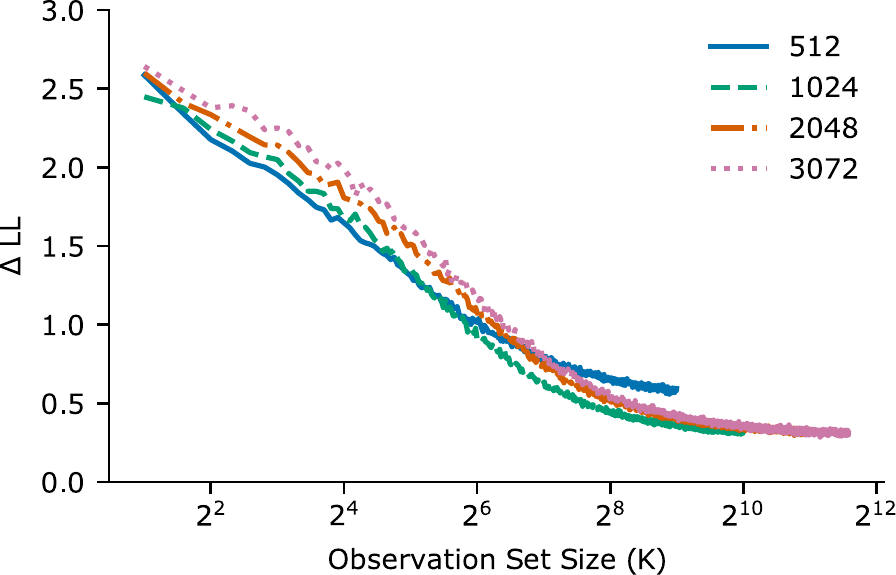}
  \caption{Predictive accuracy versus ground truth. Legend indicates the maximum observation set used during training. See \nameref{section:predictive_accuracy} for the definition of $\Delta LL$.}
  \label{fig:accuracy}
\end{figure}

\subsection{Predictive Accuracy versus observation set size}

\label{section:predictive_accuracy}
To quantify the predictive performance of this model for different numbers of observations, we computed the difference between the negative log-likelihood of the response under the $K$-shot predictive distribution to the ground truth model. This was averaged across many neurons:
\begin{equation*}
    \Delta LL_k = \left< -\log p_\theta\left(y_{k+1} | \boldsymbol x_{k+1}, O_k\right) + \log p\left(y_{k+1} | \lambda_\phi\left(\boldsymbol x_{k+1}\right)\right) \right>
\end{equation*}

We found that as as the size of the observation set increased, the predictive accuracy improved up to several hundred observations and then began to saturate. We also found that increasing the maximum set size used during training had a slight benefit in the asymptotic performance when increasing from 512 to 1024 trials, but with little benefit beyond this (Fig. \ref{fig:accuracy}). These were averaged over three different seeds, with each fit producing similar performance. 

Despite trying a number of architecture variations, we could not get the asymptotic performance to quite reach ground truth. However, it performed well, with a $\Delta LL$ of $0.4$ corresponding to a correlation coefficient between the ground truth mean response, $\lambda_\phi\left(\boldsymbol x\right)$, and the model prediction mean, $\lambda_{K=1024}(\boldsymbol x)$, of $0.8$. 

\subsection{Latent variables accurately capture  the tuning function}
\label{section:reconstruction}

\begin{figure}[b]
  \centering
  \includegraphics[width=0.8\linewidth]{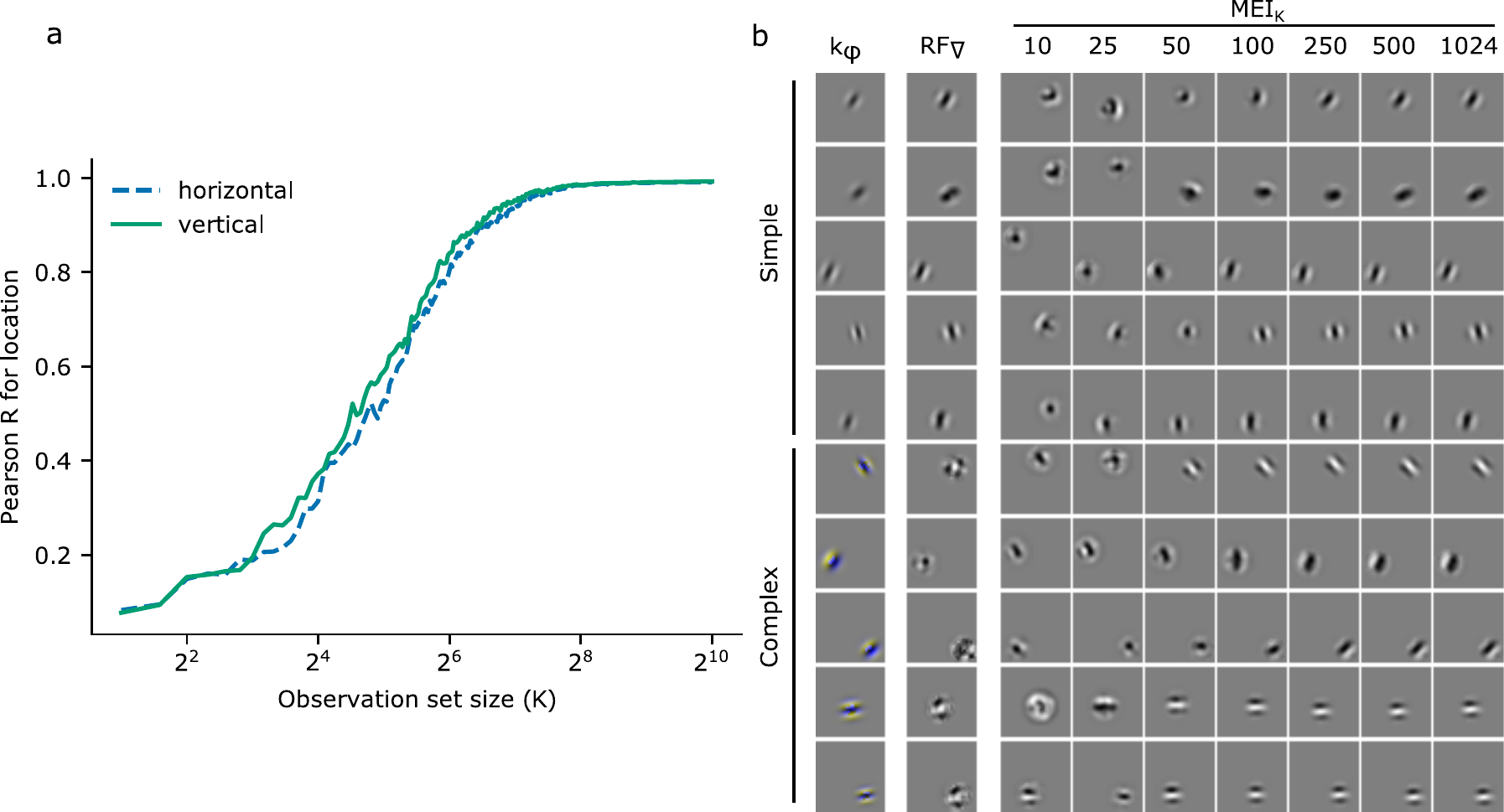}
  \caption{a) the correlation between the location latent variable, $z_k^p$, and the ground truth for increasing observations. b) Reconstruction of receptive fields (RF). Each row corresponds to a different cell with the bottom half being complex cells. The first column shows the ground truth kernels, the second column is the RF reconstructed by the gradient method, and the remaining block shows the maximally exciting images computed using increasing numbers of observations. Ground truth kernels of complex cells use pseudocolor to reflect the two phases in the energy model and any reconstruction of this energy model with the same orientation and location is equally valid, regardless of the phase.}
  \label{fig:rf_accuracy}
\end{figure}

We confirmed the information about the tuning function was correctly factorized by computing the correlation coefficient between the latent variable for location, $\boldsymbol z_k^p$, and the ground truth location of the kernel, $k_\phi$. We found that with only 64 observations there was a correlation of $0.8$, and it reached nearly $1.0$ with 256 observations (Fig.~\ref{fig:rf_accuracy}a). 

We then asked if the latent variables $(\boldsymbol z_K^p, \boldsymbol  z_K^w)$ from 1024 observations were sufficient to reconstruct the receptive field. First, we computed receptive fields as the gradient of the tuning function conditioned on the latent variables: 
\begin{equation*}
    RF_\nabla^K = \nabla_{\boldsymbol x} \left([\mathcal T \left( g_\theta(\boldsymbol x), \boldsymbol z_K^p \right), 1] \cdot u_\theta(\boldsymbol z_K^w) \right)
\end{equation*}
For simple cells the gradient showed a good correspondence to the kernel used to generate the responses, $k_\phi$ (Fig. \ref{fig:rf_accuracy}b). For complex cells it was in the correct location, but did not show the same structure as the kernel. This is expected as complex cells are not well described by a single kernel. We then computed the maximally exciting images (MEIs) for a neuron similarly to \citet{Walker2019} by maximizing the predicted response, conditioned on the latent variables sampled after an increasing number of observations:
\begin{equation*}
    MEI_K = \argmax_{\boldsymbol x} \left([\mathcal T \left( g_\theta(\boldsymbol x), \boldsymbol z_K^p \right), 1] \cdot u_\theta(\boldsymbol z_K^w) \right) - \kappa \Vert \boldsymbol x \Vert
\end{equation*}
With $\kappa=0.01$ to regularize the images.  As desired, MEIs computed with more observations converged towards the ground truth kernels, with complex cells having an anticipated random phase offset.

\subsection{Latent dimensionality and stimulus complexity}
\label{section:tuning_complexity}
\begin{figure}
  \centering
  \includegraphics[width=\linewidth]{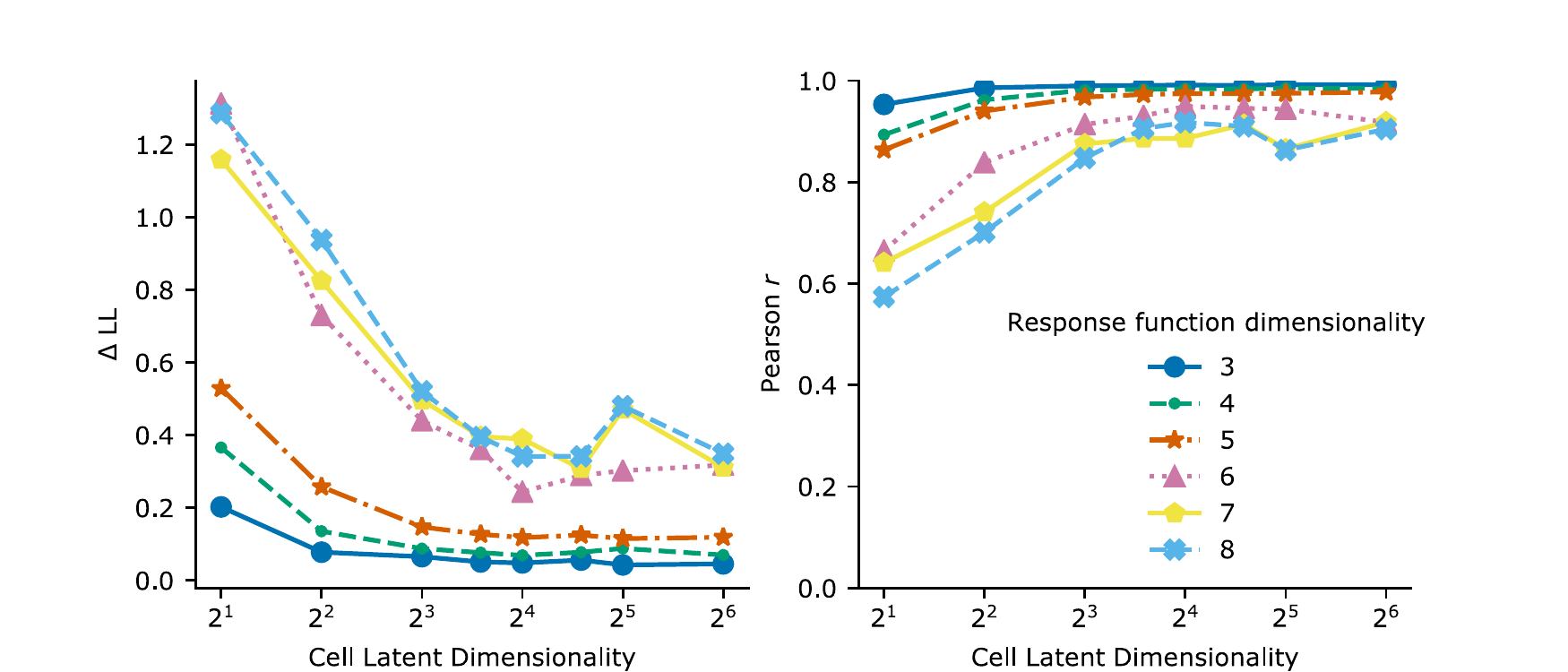}
  \caption{\textbf{Left}: The difference between the predictive log likelihood and ground truth as the dimension of the tuning function properties latent variable increases. Each line reflects an increasingly complex RF space. \textbf{Right}: The correlation between the ground truth mean firing rate and the predicted mean firing rate for the same data.}
  \label{fig:latent_dimension}
\end{figure}

We also studied how important the tuning function property's latent dimension $D$, with $\boldsymbol z_K^w \in \mathbb R^D$, was to the predictive performance by increasing it from 2 to 64 (all experiments above used 64). We did this with different complexities of the simulated receptive fields by reducing the number of parameters in $\phi$ that were randomized. In all experiments the orientation and location of the tuning function was randomized ($\phi \in \mathbb R^3$). We increased the tuning function dimensions by then additionally randomizing (in order): frequency, width, phase offset, simple only versus simple and complex cells, and scale. Because this analysis involved refitting many models, we performed it with $16\times 16$ stimuli. We found the performance improved with greater model capacity (tuning function properties latent dimension) and this impact was much more pronounced for more complex (higher dimensional) tuning functions (Fig.~\ref{fig:latent_dimension}). Randomizing the phase offset produced the greatest reduction in predictive accuracy, although performance still remained quite good with high correlations between the model predictions and ground truth. Encouragingly, including complex cells did not produce a significant change in performance.

\section{Experiments with real neural responses}

We next tested our approach on real visual responses recorded with the same experimental paradigm as in \citet{Walker2019}, and found it had a comparable predictive performance to optimization-based approaches. The data consists of pairs of neural population responses and grayscale visual stimuli sampled and cropped from ImageNet, isotropically downsampled to $64\times 36$\,px, with a resolution of $0.53$\,ppd (pixels per degree of visual angle). The neural responses were recorded from layer L2/3 of the primary visual cortex (area V1) of the mouse, using a wide field two photon microscope. A single scan contained the responses of approximately 5000--9000 neurons to up to 6000 images. 

\begin{figure}
  \begin{center}

  \includegraphics[width=0.5\linewidth]{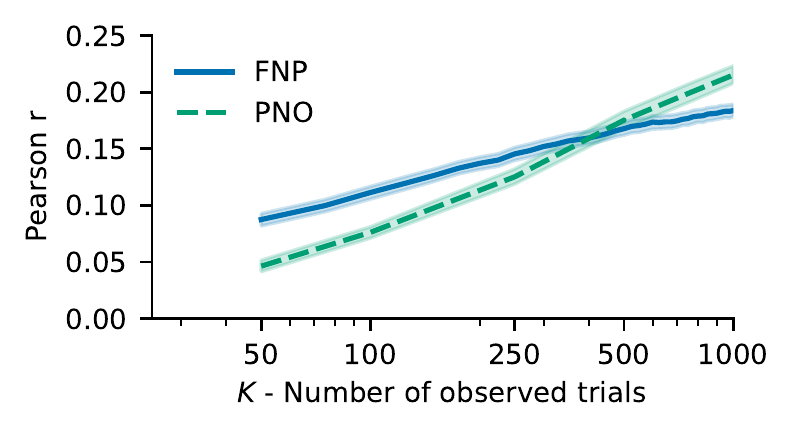}

  \end{center}
  \caption{Performance of a FNP for $K$-shot prediction for new neurons compared a traditional approach with per-neuron optimization (PNO) for $K$ up to 1000 trials}
  \label{fig:real_predictions}
\end{figure}

We trained an FNP on 57,533 mouse V1 neurons collected across 19 different scans and tested it on 1000 neurons from a hold-out scan (i.e. never seen during training). 
During testing, we assigned the latent variables assigned to their mean values: $\boldsymbol z_K^p:=\boldsymbol \mu(s_K^p)$ and $\boldsymbol z_K^w:=\boldsymbol \mu_K^w$, and used these in Eq.~\ref{eq:conditional_decoder}. 
We measured the $K$-shot predictive accuracy for each neuron as the correlation between the predicted mean from the conditional decoder, $\lambda(\boldsymbol x_t, \boldsymbol z_K^p, \boldsymbol z_K^w)$, and the real responses, $y_t$, for the remaining trials. 
In agreement with synthetic data, the predictive accuracy improves rapidly with the first several hundred trials and continues to improve with additional observations (Fig.~\ref{fig:real_predictions}). 
We compared the performance of our FNP to an optimization based approach similar to \citet{Klindt2017}, adapted for mouse V1, which we reference as Per Neuron Optimization (PNO). We measured the predictive performance of PNO similarly to FNP, on the same 1000 neurons with the readout optimized with $K$ trials and used to predict the response to the remaining stimuli. 
Excitingly, FNP performs well and with 1k images is almost as accurate as PNO (which is optimized for those individual cells), and even \emph{outperforms} it for smaller numbers of observations (Fig.~\ref{fig:real_predictions}). This likely arises because the FNP learns the prior distribution over tuning functions, which has a greater influence with less data. Please see the Appendix for details of both FNP and PNO fitting and testing.

These experiments also demonstrated the speed improvements for inferring the tuning function of a newly recorded neurons that FNP was designed for. While fitting the FNP to the training data took 6 days using two V100 GPUs, computing the latent variables for one thousand neurons with $K=1000$ took only 250~ms on a 1080Ti. This is in comparison to PNO which takes from 20~s to compute the readout using a pretrained CNN (Supplementary Table~\ref{table:times}). Thus an FNP is two orders of magnitude faster, enabling real-time inference of tuning functions within the time of a single stimulus presentation.

\section{Discussion}

Using a Factorized Neural Process, we are able to learn a distribution over tuning functions that can be conditioned on a set of observed stimulus-response pairs and predict the response to novel stimuli. We first focused on simulated data from simple and complex cells where we could compare the inferred tuning functions to the ground truth. Importantly, the model performed equally well when including complex cells, which is not possible for classical techniques like spike-triggered average that similarly accumulate sufficient statistics. The fact that the asymptotic log likelihood for predictions did not reach the ground truth also indicates there is room to increase the model capacity, although the correlation between the ground truth and model predictions exceeded $0.8$. 
Following prior work \cite{Klindt2017, Ecker2019, Walker2019,Sinz2018}, we restricted ourselves to a decoder that was a factorized linear readout on output of $g_\theta$, but learning a more powerful decoder could also improve the capacity.  
We then tested our approach on data from the mouse primary visual cortex in response to natural images. We found the trained FNP predicted the responses to test data with comparable accuracy as a model specifically optimized for those neurons, and even exceeded the performance when conditioned on less than 500 trials. Additionally, the FNP made these predictions orders of magnitudes more quickly than an optimization-based approach, thus opening the door to real-time, closed-loop inference of tuning functions updated after every stimulus presentation.
 
This work was motivated by real-time experimentation, but during an experiment the best way to know how a neuron responds to a stimulus is to measure it. The real need is using the observations to rapidly generate stimuli to test a hypothesis. We envision combining a FNP for rapid inference with a generator network that takes the latent representations as input and is trained \textit{in silico} prior to experiments to generate stimuli to illicit a maximal responses or reduce the uncertainty in the latent representations. We believe this general approach of training a more powerful model prior to experiments that is capable of rapid, real-time inference will be a powerful tool for the neuroscience community, and that this approach using FNPs will facilitate it.

\section*{Broader Impact}

We hope this approach will be useful to the Neuroscience community and that Factorized Neural Processes may have even broader applications for modeling functions. The ability to perform real-time, closed-loop experiments and to performances inferences with less data may reduce the amount of time to record from animals or the number of experimental sessions. 
We do not believe this methodology or the demonstrated application will disadvantage anyone.

\begin{ack}
RJC thanks the Research Accelerator Program of the Shirley Ryan AbilityLab for support during residency.
FHS is supported by the Carl-Zeiss-Stiftung and acknowledges the support of the DFG Cluster of Excellence “Machine Learning – New Perspectives for Science”, EXC 2064/1, project number 390727645.
Supported by the Intelligence Advanced Research Projects Activity (IARPA) via Department of Interior/Interior Business Center (DoI/IBC) contract number D16PC00003. The U.S. Government is authorized to reproduce and distribute reprints for Governmental purposes notwithstanding any copyright annotation thereon. Disclaimer: The views and conclusions contained herein are those of the authors and should not be interpreted as necessarily representing the official policies or endorsements, either expressed or implied, of IARPA, DoI/IBC, or the U.S. Government. \end{ack}

\small

\bibliography{references}

\begin{thebibliography}{38}
\providecommand{\natexlab}[1]{#1}
\providecommand{\url}[1]{\texttt{#1}}
\expandafter\ifx\csname urlstyle\endcsname\relax
  \providecommand{\doi}[1]{doi: #1}\else
  \providecommand{\doi}{doi: \begingroup \urlstyle{rm}\Url}\fi

\bibitem[Hubel and Wiesel(1962)]{hubel1962receptive}
D.~Hubel and T.~Wiesel.
\newblock Receptive fields, binocular interaction, and functional architecture
  in the cat's visual cortex.
\newblock \emph{Journal of Physiology}, 1962.

\bibitem[Antol{\'{i}}k et~al.(2016)Antol{\'{i}}k, Hofer, Bednar, and
  Mrsic-Flogel]{Antolik2016}
J{\'{a}}n Antol{\'{i}}k, Sonja~B. Hofer, James~A. Bednar, and Thomas~D.
  Mrsic-Flogel.
\newblock {Model Constrained by Visual Hierarchy Improves Prediction of Neural
  Responses to Natural Scenes}.
\newblock \emph{PLOS Computational Biology}, 12\penalty0 (6), 2016.

\bibitem[Yamins and DiCarlo(2016)]{Yamins2016}
Daniel~L.K. Yamins and James~J. DiCarlo.
\newblock {Using goal-driven deep learning models to understand sensory
  cortex}.
\newblock \emph{Nature Neuroscience}, 2016.

\bibitem[Klindt et~al.(2017)Klindt, Ecker, Euler, and Bethge]{Klindt2017}
David~A. Klindt, Alexander~S. Ecker, Thomas Euler, and Matthias Bethge.
\newblock {Neural system identification for large populations separating "what"
  and "where"}.
\newblock In \emph{Advances in Neural Information Processing Systems}. Neural
  information processing systems foundation, 2017.

\bibitem[Cadena et~al.(2019)Cadena, Denfield, Walker, Gatys, Tolias, Bethge,
  and Ecker]{Cadena2019}
Santiago~A. Cadena, George~H. Denfield, Edgar~Y. Walker, Leon~A. Gatys,
  Andreas~S. Tolias, Matthias Bethge, and Alexander~S. Ecker.
\newblock {Deep convolutional models improve predictions of macaque V1
  responses to natural images}.
\newblock \emph{PLOS Computational Biology}, 15\penalty0 (4):\penalty0
  e1006897, 2019.

\bibitem[Batty et~al.(2017)Batty, Merel, Brackbill, Heitman, Sher, Litke,
  Chichilnisky, and Paninski]{Batty2016}
Eleanor Batty, Josh Merel, Nora Brackbill, Alexander Heitman, Alexander Sher,
  Alan Litke, E.J. Chichilnisky, and Liam Paninski.
\newblock {Multilayer Recurrent Network Models of Primate Retinal Ganglion Cell
  Responses}.
\newblock In \emph{ICLR 2017}, 2017.

\bibitem[Sinz et~al.(2018)Sinz, Ecker, Fahey, Walker, Cobos, Froudarakis,
  Yatsenko, Pitkow, Reimer, and Tolias]{Sinz2018}
F.~Sinz, A.~S. Ecker, P.~Fahey, E.~Walker, E~Cobos, E.~Froudarakis,
  D.~Yatsenko, X.~Pitkow, J.~Reimer, and A.~Tolias.
\newblock {Stimulus domain transfer in recurrent models for large scale
  cortical population prediction on video}.
\newblock In \emph{Advances in Neural Information Processing Systems 31}.
  Curran Associates, Inc., 2018.

\bibitem[Yamins et~al.(2014)Yamins, Hong, Cadieu, Solomon, Seibert, and
  DiCarlo]{Yamins2014}
D.~L.~K. Yamins, H.~Hong, C.~F. Cadieu, E.~A. Solomon, D.~Seibert, and J.~J.
  DiCarlo.
\newblock {Performance-optimized hierarchical models predict neural responses
  in higher visual cortex.}
\newblock \emph{Proceedings of the National Academy of Sciences of the United
  States of America}, 2014.

\bibitem[Vintch et~al.(2015)Vintch, Movshon, and Simoncelli]{Vintch2015}
B.~Vintch, J.~A. Movshon, and E.~P. Simoncelli.
\newblock {A Convolutional Subunit Model for Neuronal Responses in Macaque V1.}
\newblock \emph{The Journal of neuroscience : the official journal of the
  Society for Neuroscience}, 35\penalty0 (44):\penalty0 14829--41, 2015.

\bibitem[Ecker et~al.(2019{\natexlab{a}})Ecker, Sinz, Froudarakis, Fahey,
  Cadena, Walker, Cobos, Reimer, Tolias, and Bethge]{Ecker2018}
Alexander~S. Ecker, Fabian~H. Sinz, Emmanouil Froudarakis, Paul~G. Fahey,
  Santiago~A. Cadena, Edgar~Y. Walker, Erick Cobos, Jacob Reimer, Andreas~S.
  Tolias, and Matthias Bethge.
\newblock A rotation-equivariant convolutional neural network model of primary
  visual cortex.
\newblock In \emph{International Conference on Learning Representations, ICLR},
  2019{\natexlab{a}}.

\bibitem[Walker et~al.(2019)Walker, Sinz, Cobos, Muhammad, Froudarakis, Fahey,
  Ecker, Reimer, Pitkow, and Tolias]{Walker2019}
Edgar~Y. Walker, Fabian~H. Sinz, Erick Cobos, Taliah Muhammad, Emmanouil
  Froudarakis, Paul~G. Fahey, Alexander~S. Ecker, Jacob Reimer, Xaq Pitkow, and
  Andreas~S. Tolias.
\newblock {Inception loops discover what excites neurons most using deep
  predictive models}.
\newblock \emph{Nature Neuroscience}, 2019.

\bibitem[Bashivan et~al.(2019)Bashivan, Kar, and DiCarlo]{Bashivan2019}
Pouya Bashivan, Kohitij Kar, and James~J. DiCarlo.
\newblock {Neural population control via deep image synthesis}.
\newblock \emph{Science}, 2019.

\bibitem[Garnelo et~al.(2018{\natexlab{a}})Garnelo, Schwarz, Rosenbaum, Viola,
  Rezende, Eslami, and Teh]{Garnelo2018a}
Marta Garnelo, Jonathan Schwarz, Dan Rosenbaum, Fabio Viola, Danilo~J. Rezende,
  S.~M.~Ali Eslami, and Yee~Whye Teh.
\newblock {Neural Processes}.
\newblock 2018{\natexlab{a}}.

\bibitem[Garnelo et~al.(2018{\natexlab{b}})Garnelo, Rosenbaum, Maddison,
  Ramalho, Saxton, Shanahan, Teh, Rezende, and Eslami]{Garnelo2018}
Marta Garnelo, Dan Rosenbaum, Christopher Maddison, Tiago Ramalho, David
  Saxton, Murray Shanahan, Yee~Whye Teh, Danilo Rezende, and S.~M.~Ali Eslami.
\newblock Conditional neural processes.
\newblock In \emph{Proceedings of the 35th International Conference on Machine
  Learning}, 2018{\natexlab{b}}.

\bibitem[Kim et~al.(2019)Kim, Mnih, Schwarz, Garnelo, Eslami, Rosenbaum,
  Vinyals, and Teh]{Kim2019}
Hyunjik Kim, Andriy Mnih, Jonathan Schwarz, Marta Garnelo, Ali Eslami, Dan
  Rosenbaum, Oriol Vinyals, and Yee~Whye Teh.
\newblock Attentive neural processes.
\newblock In \emph{International Conference on Learning Representations, ICLR},
  2019.

\bibitem[Zaheer et~al.(2017)Zaheer, Kottur, Ravanbakhsh, Poczos, Salakhutdinov,
  and Smola]{Zaheer2017}
Manzil Zaheer, Satwik Kottur, Siamak Ravanbakhsh, Barnabas Poczos, Russ~R
  Salakhutdinov, and Alexander~J Smola.
\newblock Deep sets.
\newblock In I.~Guyon, U.~V. Luxburg, S.~Bengio, H.~Wallach, R.~Fergus,
  S.~Vishwanathan, and R.~Garnett, editors, \emph{Advances in Neural
  Information Processing Systems 30}, pages 3391--3401. Curran Associates,
  Inc., 2017.

\bibitem[Cohen and Welling(2016)]{Cohen2016}
Taco~S. Cohen and Max Welling.
\newblock {Group Equivariant Convolutional Networks}.
\newblock \emph{33rd International Conference on Machine Learning, ICML}, 2016.

\bibitem[Cohen and Welling(2017)]{Cohen2016a}
Taco~S. Cohen and Max Welling.
\newblock Steerable cnns.
\newblock In \emph{5th International Conference on Learning Representations,
  {ICLR} 2017, Toulon, France, April 24-26, 2017, Conference Track
  Proceedings}, 2017.

\bibitem[Weiler et~al.(2018)Weiler, Hamprecht, and Storath]{Weiler2017}
Maurice Weiler, Fred~A. Hamprecht, and Martin Storath.
\newblock Learning steerable filters for rotation equivariant cnns.
\newblock In \emph{The IEEE Conference on Computer Vision and Pattern
  Recognition (CVPR)}, 2018.

\bibitem[Bekkers et~al.(2018)Bekkers, Lafarge, Veta, Eppenhof, Pluim, and
  Duits]{Bekkers2018}
Erik~J. Bekkers, Maxime~W. Lafarge, Mitko Veta, Koen A.~J. Eppenhof, Josien
  P.~W. Pluim, and Remco Duits.
\newblock Roto-translation covariant convolutional networks for medical image
  analysis.
\newblock In \emph{MICCAI}, 2018.

\bibitem[Lafarge et~al.(2020)Lafarge, Bekkers, Pluim, Duits, and
  Veta]{Lafarge2020}
Maxime~W. Lafarge, Erik~J. Bekkers, Josien P.~W. Pluim, Remco Duits, and Mitko
  Veta.
\newblock Roto-translation equivariant convolutional networks: Application to
  histopathology image analysis, 2020.

\bibitem[Ecker et~al.(2019{\natexlab{b}})Ecker, Sinz, Froudarakis, Fahey,
  Cadena, Walker, Cobos, Reimer, Tolias, and Bethge]{Ecker2019}
A.~S. Ecker, F.~H. Sinz, E.~Froudarakis, P.~G. Fahey, S.~A. Cadena, E.~Y.
  Walker, E.~Cobos, J.~Reimer, A.~S. Tolias, and M.~Bethge.
\newblock A rotation-equivariant convolutional neural network model of primary
  visual cortex.
\newblock \emph{International Conference on Learning Representations, ICLR},
  2019{\natexlab{b}}.

\bibitem[Ustyuzhaninov et~al.(2020)Ustyuzhaninov, Cadena, Froudarakis, Fahey,
  Walker, Cobos, Reimer, Sinz, Tolias, Bethge, and Ecker]{Ustyuzhaninov2019}
Ivan Ustyuzhaninov, Santiago~A. Cadena, Emmanouil Froudarakis, Paul~G. Fahey,
  Edgar~Y. Walker, Erick Cobos, Jacob Reimer, Fabian~H. Sinz, Andreas~S.
  Tolias, Matthias Bethge, and Alexander~S. Ecker.
\newblock Rotation-invariant clustering of neuronal responses in primary visual
  cortex.
\newblock In \emph{International Conference on Learning Representations, ICLR},
  2020.

\bibitem[Huang et~al.(2017)Huang, Liu, {Van Der Maaten}, and
  Weinberger]{Huang2017}
Gao Huang, Zhuang Liu, Laurens {Van Der Maaten}, and Kilian~Q. Weinberger.
\newblock {Densely connected convolutional networks}.
\newblock In \emph{Proceedings - 30th IEEE Conference on Computer Vision and
  Pattern Recognition, CVPR}, 2017.

\bibitem[Veeling et~al.(2018)Veeling, Linmans, Winkens, Cohen, and
  Welling]{Veeling2018}
Bastiaan~S. Veeling, Jasper Linmans, Jim Winkens, Taco Cohen, and Max Welling.
\newblock Rotation equivariant cnns for digital pathology.
\newblock In \emph{Medical Image Computing and Computer Assisted Intervention,
  MICCAI}, 2018.

\bibitem[Ioffe and Szegedy(2015)]{Ioffe2015}
Sergey Ioffe and Christian Szegedy.
\newblock {Batch normalization: Accelerating deep network training by reducing
  internal covariate shift}.
\newblock In \emph{32nd International Conference on Machine Learning, ICML},
  2015.

\bibitem[Szegedy et~al.(2015)Szegedy, Liu, Jia, Sermanet, Reed, Anguelov,
  Erhan, Vanhoucke, and Rabinovich]{Szegedy2015}
Christian Szegedy, Wei Liu, Yangqing Jia, Pierre Sermanet, Scott Reed, Dragomir
  Anguelov, Dumitru Erhan, Vincent Vanhoucke, and Andrew Rabinovich.
\newblock {Going deeper with convolutions}.
\newblock In \emph{Proceedings of the IEEE Computer Society Conference on
  Computer Vision and Pattern Recognition, CVPR}, 2015.

\bibitem[Lin et~al.(2013)Lin, Chen, and Yan]{Lin2013}
Min Lin, Qiang Chen, and Shuicheng Yan.
\newblock {Network In Network (paper)}.
\newblock \emph{arXiv preprint}, page~10, 2013.

\bibitem[Clevert et~al.(2016)Clevert, Unterthiner, and Hochreiter]{Clevert2015}
Djork-Arn{\'{e}} Clevert, Thomas Unterthiner, and Sepp Hochreiter.
\newblock {Fast and Accurate Deep Network Learning by Exponential Linear Units
  (ELUs)}.
\newblock \emph{4th International Conference on Learning Representations,
  ICLR}, 2016.

\bibitem[Jaderberg et~al.(2015)Jaderberg, Simonyan, Zisserman, and
  kavukcuoglu]{jaderberg2016spatial}
Max Jaderberg, Karen Simonyan, Andrew Zisserman, and koray kavukcuoglu.
\newblock Spatial transformer networks.
\newblock In C.~Cortes, N.~D. Lawrence, D.~D. Lee, M.~Sugiyama, and R.~Garnett,
  editors, \emph{Advances in Neural Information Processing Systems 28}. Curran
  Associates, Inc., 2015.

\bibitem[Kingma and Welling(2014)]{Kingma2014}
Diederik~P. Kingma and Max Welling.
\newblock {Auto-encoding variational bayes}.
\newblock In \emph{2nd International Conference on Learning Representations,
  ICLR}, 2014.

\bibitem[Yatsenko et~al.(2015)Yatsenko, Reimer, Ecker, Walker, Sinz, Berens,
  Hoenselaar, James~Cotton, Siapas, and Tolias]{Yatsenko2015}
Dimitri Yatsenko, Jacob Reimer, Alexander~S. Ecker, Edgar~Y. Walker, Fabian
  Sinz, Philipp Berens, Andreas Hoenselaar, R.~James~Cotton, Athanassios~S.
  Siapas, and Andreas~S. Tolias.
\newblock Datajoint: managing big scientific data using matlab or python.
\newblock \emph{bioRxiv}, 2015.

\bibitem[Russakovsky et~al.(2015)Russakovsky, Deng, Su, Krause, Satheesh, Ma,
  Huang, Karpathy, Khosla, Bernstein, Berg, and Fei-Fei]{Russakovsky2015}
Olga Russakovsky, Jia Deng, Hao Su, Jonathan Krause, Sanjeev Satheesh, Sean Ma,
  Zhiheng Huang, Andrej Karpathy, Aditya Khosla, Michael Bernstein,
  Alexander~C. Berg, and Li~Fei-Fei.
\newblock {ImageNet Large Scale Visual Recognition Challenge}.
\newblock \emph{International Journal of Computer Vision}, 2015.

\bibitem[Chrabaszcz et~al.(2017)Chrabaszcz, Loshchilov, and
  Hutter]{Chrabaszcz2017}
Patryk Chrabaszcz, Ilya Loshchilov, and Frank Hutter.
\newblock A downsampled variant of imagenet as an alternative to the cifar
  datasets.
\newblock \emph{arXiv preprint arXiv:1707.08819}, 2017.

\bibitem[van~der Walt et~al.(2014)van~der Walt, {S}ch\"onberger,
  {Nunez-Iglesias}, {B}oulogne, {W}arner, {Y}ager, {G}ouillart, {Y}u, and the
  scikit-image contributors]{scikit-image}
{S}t\'efan van~der Walt, {J}ohannes~{L}. {S}ch\"onberger, {J}uan
  {Nunez-Iglesias}, {F}ran\c{c}ois {B}oulogne, {J}oshua~{D}. {W}arner, {N}eil
  {Y}ager, {E}mmanuelle {G}ouillart, {T}ony {Y}u, and the scikit-image
  contributors.
\newblock scikit-image: image processing in {P}ython.
\newblock \emph{PeerJ}, 2014.

\bibitem[Abadi et~al.(2015)Abadi, Agarwal, Barham, Brevdo, Chen, Citro,
  Corrado, Davis, Dean, Devin, Ghemawat, Goodfellow, Harp, Irving, Isard, Jia,
  Jozefowicz, Kaiser, Kudlur, Levenberg, Man\'{e}, Monga, Moore, Murray, Olah,
  Schuster, Shlens, Steiner, Sutskever, Talwar, Tucker, Vanhoucke, Vasudevan,
  Vi\'{e}gas, Vinyals, Warden, Wattenberg, Wicke, Yu, and
  Zheng]{tensorflow2015-whitepaper}
Mart\'{\i}n Abadi, Ashish Agarwal, Paul Barham, Eugene Brevdo, Zhifeng Chen,
  Craig Citro, Greg~S. Corrado, Andy Davis, Jeffrey Dean, Matthieu Devin,
  Sanjay Ghemawat, Ian Goodfellow, Andrew Harp, Geoffrey Irving, Michael Isard,
  Yangqing Jia, Rafal Jozefowicz, Lukasz Kaiser, Manjunath Kudlur, Josh
  Levenberg, Dandelion Man\'{e}, Rajat Monga, Sherry Moore, Derek Murray, Chris
  Olah, Mike Schuster, Jonathon Shlens, Benoit Steiner, Ilya Sutskever, Kunal
  Talwar, Paul Tucker, Vincent Vanhoucke, Vijay Vasudevan, Fernanda Vi\'{e}gas,
  Oriol Vinyals, Pete Warden, Martin Wattenberg, Martin Wicke, Yuan Yu, and
  Xiaoqiang Zheng.
\newblock {TensorFlow}: Large-scale machine learning on heterogeneous systems,
  2015.

\bibitem[Kingma and Ba(2015)]{Kingma2015}
Diederik~P. Kingma and Jimmy~Lei Ba.
\newblock {Adam: A method for stochastic optimization}.
\newblock In \emph{3rd International Conference on Learning Representations,
  ICLR}, 2015.

\bibitem[Chollet(2017)]{Chollet2017}
François Chollet.
\newblock {Xception: Deep learning with depthwise separable convolutions}.
\newblock In \emph{Proceedings - 30th IEEE Conference on Computer Vision and
  Pattern Recognition, CVPR}, 2017.

\end{thebibliography}

\cleardoublepage
\section{Appendix}
\subsection{Simulated Responses Tests}

\subsubsection{Architecture}

Our G-CNN DenseNet \cite{Huang2017} architecture had a lifting convolution layer into the group space with 32 channels and spatial kernel size $11\!\times\! 11$, followed by three Dense Blocks each with a growth factor of $k=8$ with the $1\!\times\! 1$ convolution having $4k$ channels and kernel sizes $5\!\times\! 5$, followed by two $3\!\times\! 3$ layers. Group convolutions were sampled at 8 equally spaced orientations computed with bilinear interpolation and a circular mask \cite{Bekkers2018,Lafarge2020}. The group space had a final batch normalization \cite{Ioffe2015} and ELU activation \cite{Clevert2015} and then was projected into image space by flattening the orientation dimension into the feature channels. A final $1\!\times\! 1$ convolution reduced the features to 64 channels, which we refer to as $C$. Our G-CNN implementation was based on the implementation described in \citet{Bekkers2018} found at \url{https://github.com/tueimage/SE2CNN/tree/master/se2cnn}.


The observation embedding for the location prediction, $h^p_\theta$, was a $1\!\times\! 1$ convolution with three layers, the first two with the same depth as the G-CNN output and the final layer with only one channel. The observation embedding for the tuning function properties latent variable, $h_\theta^w$ and $u_\theta$ were both two layer MLPs networks each with with $C+1$ channels and no activation function after the final layer of $u_\theta$. The activation function throughout was $\texttt{ELU}$.

\subsubsection{Training and Testing}

The simulated training data consisted of 5000 neuron parameters used to generate receptive fields and responses as described in Section~\ref{section:simulated_responses}. The linear component of the receptive field was generated with scikit-image \texttt{gabor\_kernel} \cite{scikit-image}. The parameters were selected randomly or fixed for the experiments described in Section~\ref{section:tuning_complexity}. The set of values used for both the training and testing data are shown in Table~\ref{table:rf_params}. Stimuli were selected from a fixed 10k images from the training split of ImageNet. The testing performance was measured using images from the testing split of ImageNet using unique tuning function parameter samples.

The experiments in Section~\ref{section:reconstruction} used image sizes of $32 \times 32$. For the experiments in Sections \ref{section:predictive_accuracy} and \ref{section:tuning_complexity} the stimuli were downsampled to $16 \times 16$ due to the number of models trained.

\begin{table}
\centering
\begin{tabular}{ || l | l | l || }\hline
 \textbf{Parameter ($\phi_i$)} & Distribution & Constant \\
 \hline
 X and Y location & Uniform spanning inner $2/3$ of range & NA \\ 
 \hline
 Orientation & Uniform[$0$, $\pi$] & NA \\
 \hline
 Frequency & $0.08 + 0.16 / (1 + e^{\mathcal N(0,1)})$ & 0.16\\
 \hline
 Width & $2 + 1 / (1 + e^{\mathcal N(0,1)})$  & 2\\
 \hline
 Phase & Uniform[$0$, $\pi$] & 0\\
 \hline
 Complex? & Bernoulli[$0.5$] & 0 \\
 \hline
 Scale & $1.0 + 0.5 \cdot \tanh ( \mathcal N(0, 1)$ ) & 1\\
 \hline
\end{tabular}
\vspace{1em}
\caption{Parameters used for simulated receptive fields}
\label {table:rf_params}
\end{table}

The model was implemented in TensorFlow 2.0 \cite{tensorflow2015-whitepaper} with TensorFlow Probability used for all distributions. We used the ADAM optimizer \cite{Kingma2015} with learning rate $10^{-3}$ with L2 weight decay $10^{-6}$ based on prior hyperparameter optimization. We used early stopping when the training loss did not improve for 75 steps. Models were fit on a Titan RTX, although for larger stimulus size and models we would distribute across multiple GPUs. To support multiple-GPU training, we created mini-batches of (number of GPU x trials x neurons per GPU) size. This was because TensorFlow treats the first dimension as the batch dimension and splits along that when distributing across GPUs, and so if the trial dimension was the first one the data had a shorter maximum length. Depending on the complexity of the tuning functions and model architecture, fitting took from several hours to several days. Experiments including different architectures and hyperparameters were managed using DataJoint \cite{Yatsenko2015}. 

\subsection{V1 Responses FNP Architecture and training details}


The architecture used for fitting the \emph{in vivo} data was the same as just described for the simulated responses, with the following changes. The G-CNN had a lifting convolution layer into the group space with 16 channels and spatial size of $17 \times 17$, followed by three G-CNN Dense Blocks: the first with spatial dimension $5 \times 5$ with 8 channels, followed by two of size $3 \times 3$ with four channels (resulting in a total of 32 group channels after these four layers). The final $1\!\times\! 1$ convolution reduced to 16 channels. The latent variable had dimension of 24. These parameters were changed primarily to reduce the memory during training while maintaining enough capacity for good predictions. Because these experiments used images of $36 \times 64$ they used significantly more memory than the $16 \times 16$ and $32 \times 32$ images used in simulations. Similarly the batch size was limited to only several neurons because each ``sample'' in a batch is the entire set of $K$ stimuli and responses. The FNP was trained for six days on two V100 GPUs using a batch size of 4 neurons per GPU and a total size of 1024 stimuli. 

Neurons and trials were randomly selected from the 57,533 V1 neurons. Each batch used neurons from a randomly selected one of the 19 training scans, due to the need to have responses to the same stimuli. Because the aggregation operation is order invariant, permuting the order \emph{within} the observation batch does not change the inferred latent distribution. However, the ordering was randomized to place individual trials at random $K$ positions for the $K$-shot prediction. The optimizer, learning rate and weight decay were unchanged from our experiments with simulated data.

We also note that the Poisson likelihood is strictly not correct, because the denconvolved fluorescence traces are real while the Poisson distribution is defined on integers. However, following previous work we still used it because it seems to yield better performance in practice \cite{Klindt2017, Ecker2019, Walker2019,Sinz2018,Cadena2019}.

\subsection{PNO model fitting and testing}

We trained it on the responses of a single scan with 4,335 neurons to 5k natural scenes. We used a 4-layer convolutional core with 64 channels in each layers. In the first layer, we used a full convolution with kernel size $15 \times 15$. In the subsequent layers, we use depth separable convolutions with kernel size $13 \times 13$ \cite{Chollet2017}. We used a ELU nonlinearity \citep{Clevert2015} and batch normalization before them \citep{Ioffe2015}. We used the same readout as \citet{Klindt2017} and as the FNthat factorizes the linear mapping from the final tensor to the neural response into a spatial and a channel component. The final linear mapping was followed by an ELU nonlinarity and an offset of one, to keep the responses consistent. We used L2 regularization on the Laplace filtered input kernels (regularization coefficient $\gamma=1$), and L1 regularization on the readout coefficients (regularization coefficient $\gamma=0.0024$). The network was trained using Poisson loss and the ADAM optimizer \citep{Kingma2015} with an initial learning rate of $0.00414$. We used early stopping based on validation correlation between the model prediction and single trial neural responses using a patience of five (maximum number of steps the validation score is allowed to not improve). Every time, the patience limit was hit, we decreased the learning rate by a factor of $0.3$. We repeated this for a total of four times. At the end, we restored the best model over the training run. We trained the full network on the full dataset. Subsequently, we froze the core CNN and fit the readout to 1,000 new neurons on up to 1,000 images and measure the prediction accuracy. We used the same training protocol as above. 

\begin{table}
\centering
  \begin{tabular}{ || l | l || }\hline
 \textbf{Network} & \textbf{Time} \\
 \hline
 FNP Optimization (once) & 6 days on dual V100 \\ 
 \hline
 FNP Inference for 1000 trials & \textbf{250~ms} (1080Ti) \\
 \hline
 PNO Readout only& $\sim20$~s (1080) \\
 \hline
 PNO CNN$+$Readout & $\sim5$~min (1080) \\
 \hline
 PNO + Hyperparameters & $\sim 12$~h (1080) \\
 \hline
\end{tabular}
\vspace{1em}
\label{table:times}
\caption{Comparison of training and inference times for FNP and PNO}
\end{table}

\end{document}